\documentclass[twocolumn]{aastex6}
\usepackage{amsmath}

\begin{document}

\newcommand{\re}{$R_\oplus$} 
\newcommand{\me}{$M_\oplus$} 
\newcommand{\reb}{$\mathbf{R_\oplus}$} 
\newcommand{\meb}{$\mathbf{M_\oplus}$}


\title{Probabilistic Constraints on the Mass and Composition of Proxima~b}


\author{Alex Bixel\altaffilmark{2}}
\author{D\'aniel Apai\altaffilmark{1,2}}
\affil{Steward Observatory \\
933 North Cherry Avenue \\
Tucson, AZ 85721, USA}

\altaffiltext{1}{Department of Planetary Science/Lunar and Planetary Laboratory, The University of Arizona, 1640 E. University Blvd., Tucson, AZ 85718,
USA}
\altaffiltext{2}{Earths in Other Solar Systems Team, NASA Nexus for Exoplanet System Science}



\begin{abstract}
Recent studies regarding the habitability, observability, and possible orbital evolution of the indirectly detected exoplanet Proxima~b have mostly assumed a planet with $M \sim 1.3$ \me, a rocky composition, and an Earth-like atmosphere or none at all. In order to assess these assumptions, we use previous studies of the radii, masses, and compositions of super-Earth exoplanets to probabilistically constrain the mass and radius of Proxima~b, assuming an isotropic inclination probability distribution. We find it is $\sim 90\%$ likely that the planet's density is consistent with a rocky composition; conversely, it is at least $10\%$ likely that the planet has a significant amount of ice or an H/He envelope. If the planet does have a rocky composition, then we find expectation values and 95\% confidence intervals of $\left<M\right>_\text{rocky} = 1.63_{-0.72}^{+1.66}$ \me\ for its mass and $\left<R\right>_\text{rocky} = 1.07_{-0.31}^{+0.38}$ \re\ for its radius.

\end{abstract}




\section{Introduction} \label{sec:introduction}
The recent radial velocity detection of a planet in the habitable zone of the nearby M dwarf Proxima Centauri (hereafter 'Proxima~b' and 'Proxima') \citep{anglada16} has spurred over a dozen theoretical papers speculating on the planet's atmosphere \citep[e.g.,][]{brugger16,goldblatt16}, habitability \citep[e.g.,][]{ribas16,turbet16}, and orbital and formation histories \citep[e.g.,][]{barnes16,coleman17} as well as prospects for a direct detection or atmospheric characterization \citep[e.g.,][]{lovis16,luger16}. As Proxima is the nearest neighbor to the solar system, it has been suggested as a target for future space missions, including those hoping to characterize its atmosphere and search for life \citep[e.g.,][]{belikov15,schwieterman16}.

In many of these studies, authors have assumed a rocky planet with a thin atmosphere or no atmosphere at all, and some have assumed a mass near or equal to the projected mass of $M\sin(i) = 1.27_{-0.17}^{+0.20}$ \me, but little has been done to assign a degree of certainty to these assumptions. Most notably, previous studies have revealed two distinct populations of exoplanets with super-Earth radii: `rocky' planets composed almost entirely of rock, iron, and silicates with at most a thin atmosphere, and `sub-Neptune' planets which must contain a significant amount of ice or a H/He envelope \citep[e.g.,][]{rogers15,weiss14}. If there is a significant probability that Proxima~b is of the latter composition, then this should be taken into account when assessing its potential habitability or observability.

In this letter, we generate posterior distributions for the mass of Proxima~b using Monte Carlo simulations of exoplanets with an isotropic distribution of inclinations, where the radii, masses, and compositions of the simulated planets are constrained by results from combined transit and radial velocity measurements of previously detected exoplanets. By comparing the posterior mass distribution to the composition of planets as a function of mass, we determine the likelihood that Proxima~b is, in fact, a rocky world with a thin (if any) atmosphere.

\section{Prior assumptions} \label{sec:prior_assumptions}
Radial velocity and transit studies of exoplanets have yielded mass and radius measurements for a statistically significant number of targets, thereby enabling the study of how the occurrence and composition of exoplanets varies with planet radii, orbital periods, and host star type. In this section, we review previous results which we will use to place stronger constraints on the mass and composition of Proxima~b.

\subsection{$\sin(i)$ distribution} \label{sini_distribution}
It can be shown \citep[e.g.,][]{ho11} that the probability distribution of $\sin(i)$ corresponding to an isotropic inclination distribution is

\begin{equation} \label{equation:sini_distribution}
P(\sin(i)) = \sin(i)/\sqrt{1-\sin^2(i)}
\end{equation}

Since this distribution peaks at $\sin(i) = 1$, the mass distribution of an RV-detected planet - assuming no prior constraints on the mass - peaks at the minimum mass $M_0$.


In their models of the possible orbital histories of Proxima b, \citet{barnes16} find that galactic tides could have inflated the eccentricity of the host star's (at the time unconfirmed) orbit around the $\alpha$ Cen binary, leading to encounters within a few hundred AU and the possible disruption of Proxima's planetary system. If so, this could affect the likely inclination of the planet in a non-isotropic way. However, \citet{kervella17} have presented radial velocity measurements showing that Proxima is gravitationally bound to the $\alpha$ Cen system with an orbital period of 550,000 years, an eccentricity of $\sim 0.5$, and a periapsis distance of 4,200 AU. At this distance, the ratio of Proxima's gravitational field to that of $\alpha$ Cen at the planet's orbit ($\sim 0.05$ AU) is greater than $10^8$; unless Proxima's orbit was significantly more eccentric in the past, it seems unlikely that $\alpha$ Cen would have disrupted the system.

\subsection{Occurrence rates for M dwarfs} \label{sec:occurrence_rates}
\begin{figure}
\centering
\textbf{Occurrence Rates for M Dwarf Planets\\(with $\mathbf{2 < P < 50}$ days)}\par\medskip
\includegraphics[width=\columnwidth]{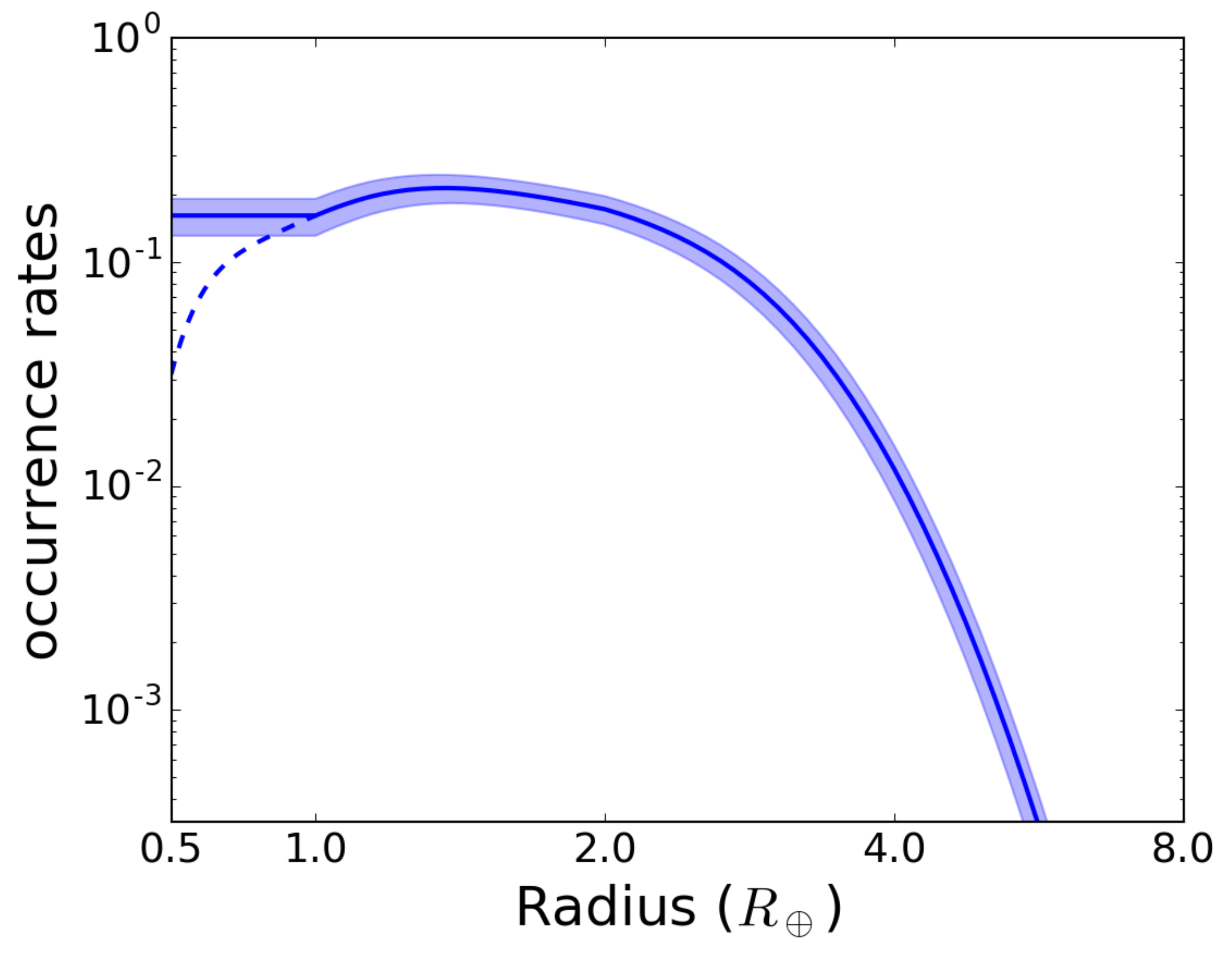}
\par\medskip
\textbf{Mass-Radius Relationships for $\mathbf{R<4}$ $\mathbf{R_\oplus}$}\par\medskip
\includegraphics[width=\columnwidth]{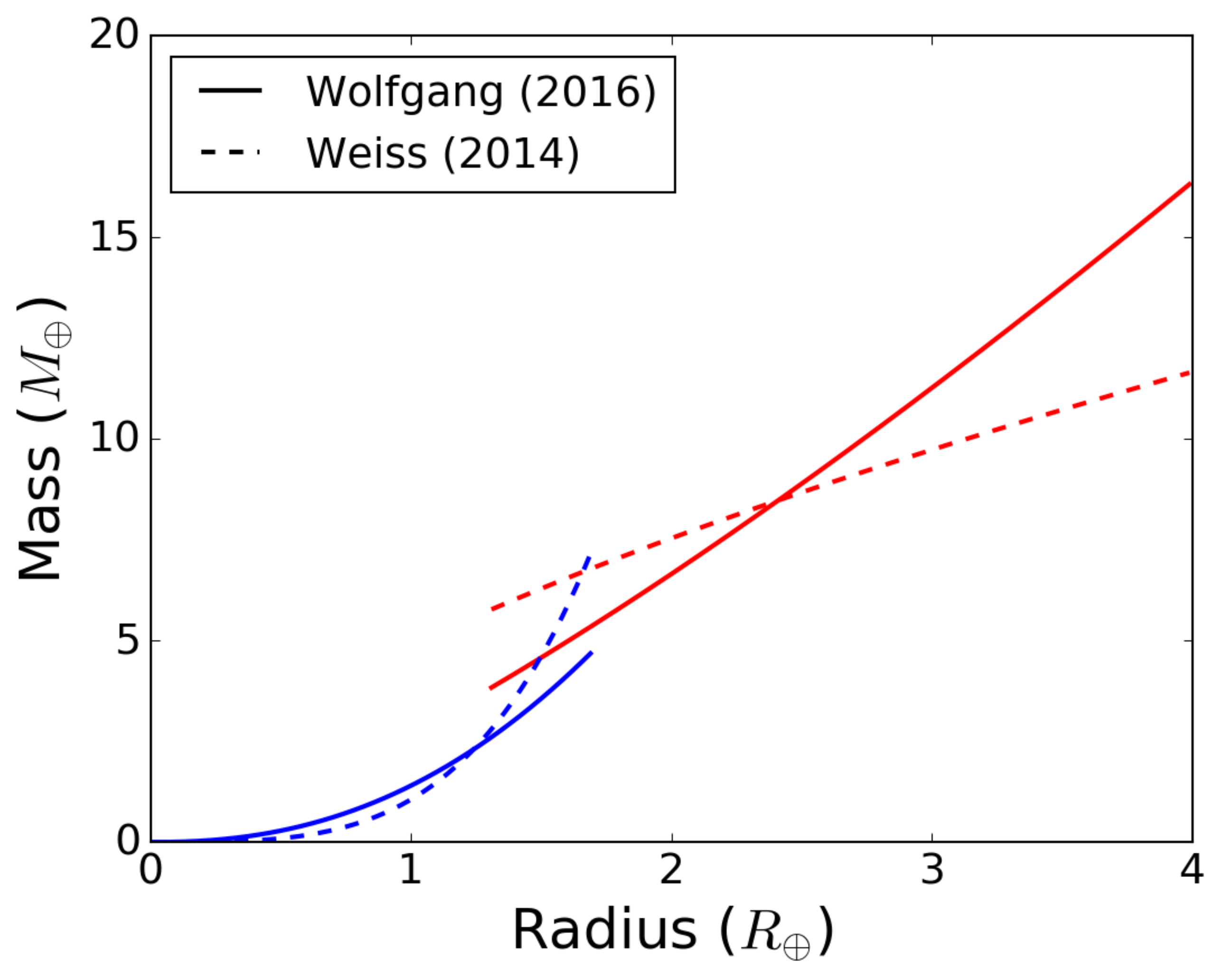}
\caption{Top: Occurrence rates from \citet{mulders15}, fitted by a regression curve. We adjust the rates below 1 \re\ (dotted) to be flat, since the sample is incomplete in this range. Bottom: Mass-radius relationships for the rocky (blue) and sub-Neptune (red) populations. The plotted relationships are from \citet{wolfgang16} (solid) and \citet{weiss14} (dashed).\label{fig:occurrence_rates} \label{fig:mass-radius_relationship}}
\end{figure}

\citet{mulders15} provide up-to-date occurrence rates of planets around M dwarf stars from the \emph{Kepler} mission. The sample is limited to $2<P<50$ days, over which they find the occurrence rates to be mostly independent of the period. The binned rates and a regression curve, as well as their uncertainties, are presented in Figure \ref{fig:occurrence_rates}.

\emph{Kepler} statistics for M dwarfs remain incomplete below 1 \re , but complete statistics for earlier-type stars suggest a flat distribution for $0.7 < R < 1$ \re\ \citep{mulders15}. Since mass-radius relationships typically find a strong dependence of mass on radius ($M \propto R^{3-4}$) \citep[e.g.][]{weiss14,wolfgang16}, we assume \emph{a priori} that Proxima~b ($M \gtrsim 1.3$ \me) is larger than $0.7$ \re . Therefore, for this letter we adopt the regression curve fitted to the binned data, but set the occurrence rates to be flat for $R < 1$ \re.

\subsection{Compositions} \label{sec:compositions}
Multiple works \citep[e.g.][]{weiss14,rogers15} have determined the existence of two distinct populations of exoplanets smaller than Neptune ($R \lesssim 4$ \re): a small radius population with densities consistent with an entirely iron and silicate composition (hereafter `rocky'), and a large radius population with lower density planets which must have significant amounts of ice or a thick H/He atmosphere (hereafter `sub-Neptunes').

\citet{rogers15} studies the abundance of planets of each composition as a function of radius. They define $f_\alpha(R)$ as the likelihood that a planet of radius $R$ is dense enough to be consistent with a rocky composition, and determine $f_\alpha$ for a sample of planets with known masses and radii. They suggest fitting the data with a two-parameter linear model:

\begin{multline} \label{equation:rogers_linear}
f_{\alpha}\left(R_P, R_{\text{thresh}}, \Delta R\right) \\
= \begin{cases} 
1 & R_P < R_{\text{thresh}} - \frac{1}{2} \Delta R \\ 
0.5 + \dfrac{R_\text{thresh}-R_P}{\Delta R} & \left|R_P-R_\text{thresh}\right| < \frac{1}{2} \Delta R\\
0 & R_P > R_{\text{thresh}} + \frac{1}{2} \Delta R  \end{cases}
\end{multline}

They find a step function to best describe the data, with $\Delta R$ fixed at zero and $R_\text{thresh} \approx 1.5$ \re. For the purposes of this letter, we prefer this fit, but will also vary $R_\text{thresh}$ and $\Delta R$ to see how they affect our results.

We stress that a planet for which $f_\alpha = 1$ is only \emph{sufficiently} dense to be rocky; we still cannot necessarily exclude an ice or volatile component. Here, we will assume that all planets for which $f_\alpha = 1$ follow the low-radius M-R relationships given in the following section, which were empirically fitted without prior knowledge of the planets' compositions. For simplicity, we refer to these as `rocky' planets, and the other population as `sub-Neptunes', but we will revisit this distinction later on.

Since Proxima b is in the habitable zone, it receives an amount of stellar flux comparable to that received by Earth, so we should bear in mind the possibility that the volatile envelope of a sub-Neptune could be lost due to photoevaporation. \citet{owen16} model rocky planets with thick H/He envelopes in the habitable zones of M dwarfs, finding that planets with $M > 0.8$ \me\ maintain their envelopes over Gyr timescales and are therefore uninhabitable. The 2$\sigma$ lower limit on the minimum mass of Proxima b is $0.93$ \me, so it is unlikely that any H/He envelope on the planet would evaporate under this rule. However, we note that this study focuses on planets with a primarily rocky composition, so it may not be directly applicable to habitable zone sub-Neptunes.

Additionally, \citet{zahnle13} empirically define boundaries for atmospheric evaporation as a function of stellar heating, escape velocity, and atmospheric composition. In particular, a planet receiving an Earth-like flux must have an escape velocity above $\sim 8$ km/s in order to maintain an H$_2$ atmosphere for 5 Gyr. We will revisit this requirement in Section \ref{sec:escape}.

\subsection{Mass-radius relationships} \label{sec:mass-radius_relationships}
Empirically derived relationships between exoplanet masses and radii rely on radial velocity (RV) or transit-timing variation (TTV) measurements of transiting exoplanet masses. \citet{weiss14} fit a mass-radius (hereafter M-R) relationship to a sample of 65 transiting exoplanets, in which they find evidence for the two populations discussed in Section \ref{sec:compositions}. Through least-squares regression, they find the densities of the rocky planets to increase linearly with planet radius:
\begin{equation} \label{equation:weiss_marcy_rocky}
\rho_P = 2.43 + 3.39 \left(\frac{R_P}{R_E} \right) \text{g cm$^{-3}$}
\end{equation}
while the RV-measured masses of sub-Neptunes increase nearly linearly with planet radius:
\begin{equation} \label{equation:weiss_marcy_rv}
\frac{M_P}{M_\oplus} = 4.87\left(\frac{R_P}{R_E}\right)^{0.63}
\end{equation}

\citet{wolfgang16} use an expanded version of this data set to fit power law M-R relationships using a more statistically robust Bayesian method. For the rocky planets, they find
\begin{equation} \label{equation:wolfgang_rocky}
\frac{M_P}{M_\oplus} = 1.4 \left(\frac{R_P}{R_E}\right)^{2.3}
\end{equation}
and for the sub-Neptunes with RV-measured masses,
\begin{equation} \label{equation:wolfgang_rv}
\frac{M_P}{M_\oplus} = 2.7 \left(\frac{R_P}{R_E}\right)^{1.3}
\end{equation}

Due to the larger sample size and more robust fitting procedure, we adopt Equations \ref{equation:wolfgang_rocky} and \ref{equation:wolfgang_rv} as our preferred M-R relationships, but for completeness we consider the \citet{weiss14} relationships as well. We find that the choice of M-R relationships has a minimal impact on our final results. Both sets of relationships are plotted in Figure \ref{fig:mass-radius_relationship}.

It is important to note that the above relationships for sub-Neptunes exclude masses measured by TTV, since TTV masses have been found to be systematically lower than RV masses. This could indicate a selection bias or systematic error in the method used, but since Proxima~b was detected through RV measurements, we believe it is proper to exclude the TTV masses either way.

It is also clear that there is a significant spread in the masses of the observed planets. \citet{wolfgang16} suggest a spread of $\pm 1.9$ \me\ for the sub-Neptune planets, which we adopt for our simulations. For rocky planets, the spread is noticeably smaller. There are too few planets to constrain this spread, but it should most likely increase with mass, so we arbitrarily define the spread to be 30\% of the calculated mass.



\section{Method} \label{sec:method}
\subsection{Simulated sample} \label{sec:simulated_sample}
The fitted occurrence rates and their uncertainties ($f \pm df$) are given in even bins in log-space. We use them to generate a random sample of radii, where the number of radii in each bin ($r_0$) is selected from a normal distribution with mean value $f(r_0)$ and standard deviation $df (r_0)$. We find that the results converge for 1,000 samples of the occurrence rates, with each sample containing $\sim 10^6$ radii.

To each radius, we assign a composition (`rocky' or `sub-Neptune') based on the model of \citet{rogers15} (Equation \ref{equation:rogers_linear}), with $R_\text{thresh} = 1.5$ \re\ and $\Delta R = 0$. We then assign a mass to each radius and composition from a Gaussian distribution with mean value $M(R)$ - calculated using our chosen M-R relationships (Equations \ref{equation:wolfgang_rocky} and \ref{equation:wolfgang_rv}) - and a standard deviation $dM$ which represents the spread. We choose a spread proportional to the calculated mass for rocky planets ($dM = 0.3\times M(R)$), but a constant spread for sub-Neptunes ($dM = 1.9$ \me). We also reject negative masses, which could in principle bias the assigned masses towards higher-than-average values - however, we find that only a negligible number of masses are rejected.

Finally, we assign a line-of-sight inclination parameter $\sin(i)$ to each planet, drawn from the isotropic probability distribution discussed in Section \ref{sini_distribution}.

\subsection{Prior and posterior probability distributions} \label{sec:prior_posterior_distributions}
The prior mass and radius distributions, $P(M)$ and $P(R)$, can be derived directly from the simulated sample. Factoring in the projected minimum mass $M_0$, the posterior distributions $P(M|M_0)$ and $P(R|M_0)$ can be calculated from Bayes' formula:
\begin{equation}
P(X|M_0) = \frac{P(M_0|X)P(X)}{P(M_0)}
\end{equation}
where $X$ represents mass or radius. Since $M_0$ is known, $P(M_0)$ is just a normalizing constant. Taking $M_0 = 1.27_{-0.17}^{+0.20}$ \me\ as the projected mass of Proxima~b and the upper limit $\sigma_{M_0} = 0.20$ \me\ as its standard deviation, we calculate for each simulated planet
\begin{equation}
P_i(M_0|X) = \exp\left(-(M_0-M_i\sin_i(i))^2/2\sigma_{M_0}^2 \right)
\end{equation}

Then $P(M_0|M)$ and $P(M_0|R)$ are the average values of $P_i(M_0|X)$ for each bin in mass or radius. The prior and posterior distributions are calculated for each sample of $10^6$ planets, and the final results are taken to be the mean result of 1,000 samples.

\subsection{Posterior compositional probability}
The prior probability that a planet in a given mass bin is rocky is equal to the number of simulated rocky planets in that bin divided by the total number of planets in the same bin. Since we want to know the likelihood that Proxima~b belongs to the `rocky' population, we multiply this prior composition probability distribution by the posterior mass distribution from the previous section and integrate over all masses.

\section{Results} \label{sec:results}

\subsection{Mass distributions} \label{sec:results_mass_distributions}

\begin{figure}
\centering
\textbf{Prior mass distribution}\par\medskip
\includegraphics[width=\columnwidth]{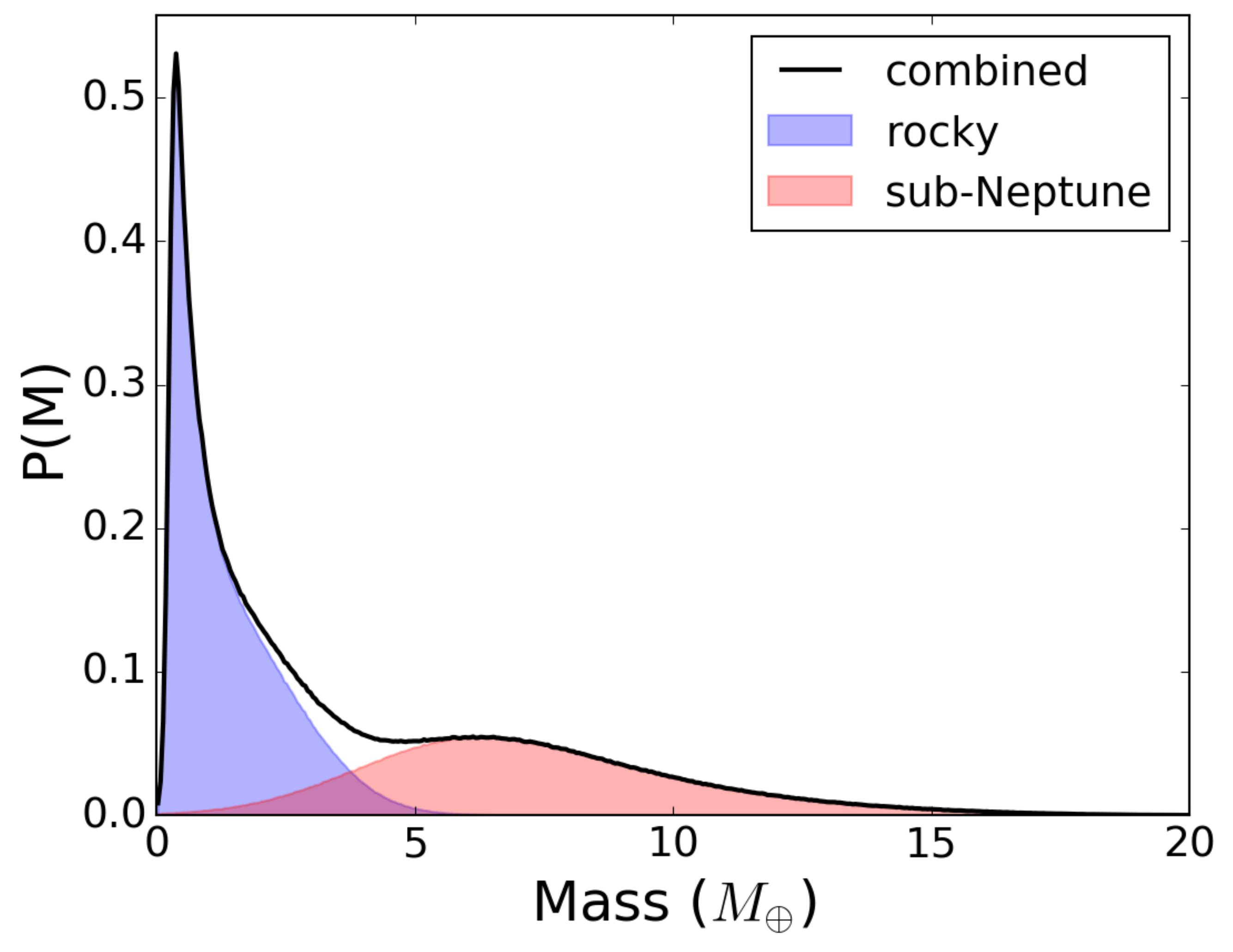}
\par\medskip
\textbf{Posterior mass distribution}\par\medskip
\includegraphics[width=\columnwidth]{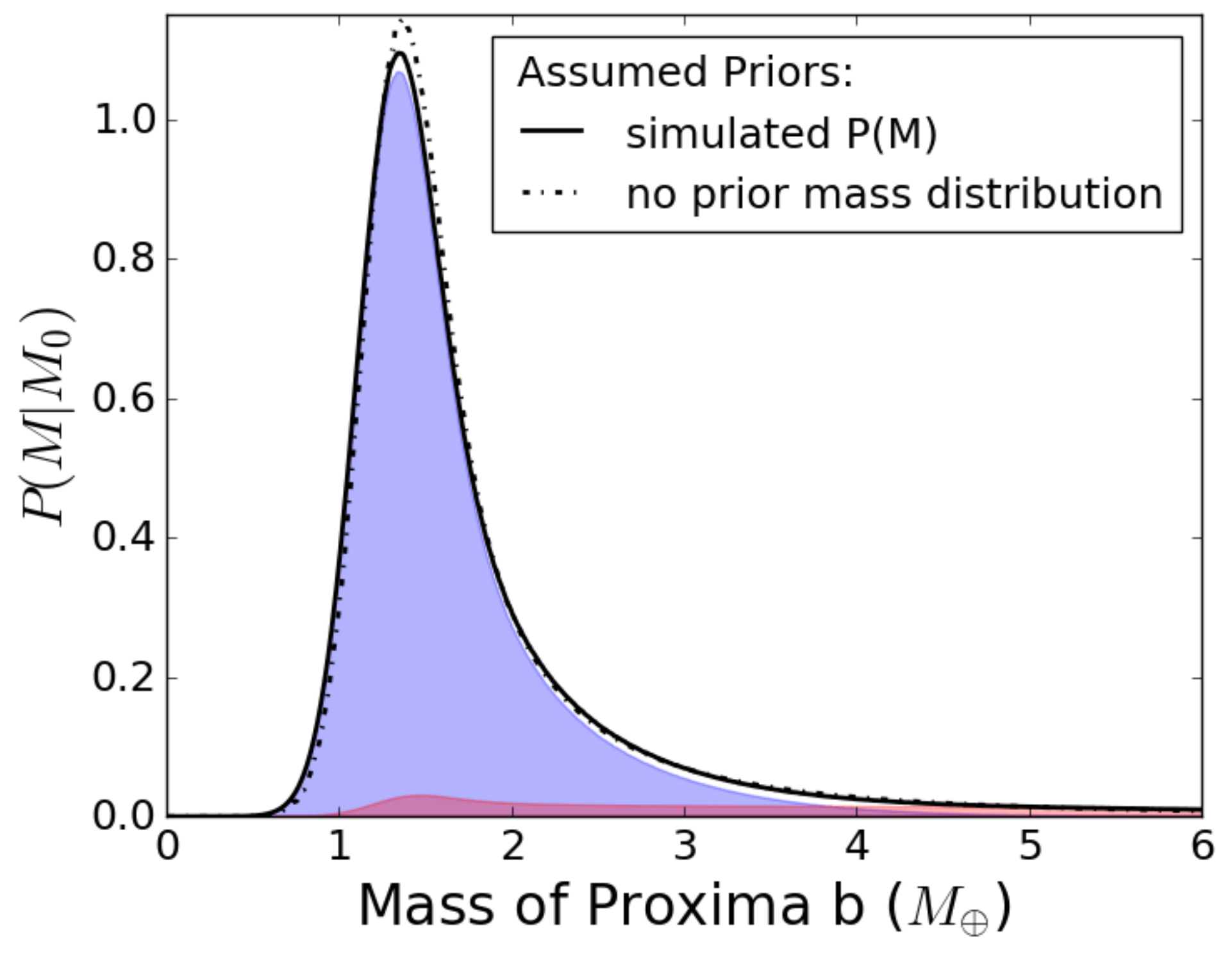}
\caption{Prior (top) and posterior (bottom) mass distributions for the simulated sample. The blue and red shaded regions represent contributions due to rocky and sub-Neptune planets, respectively. The dash-dotted line is the posterior distribution assuming a flat prior distribution. The binning is 0.01 \me. \label{fig:prior_posterior_pdf}}
\end{figure}
 
\begin{figure}
\centering
\textbf{Cumulative mass probability distribution}\par\medskip
\includegraphics[width=\columnwidth]{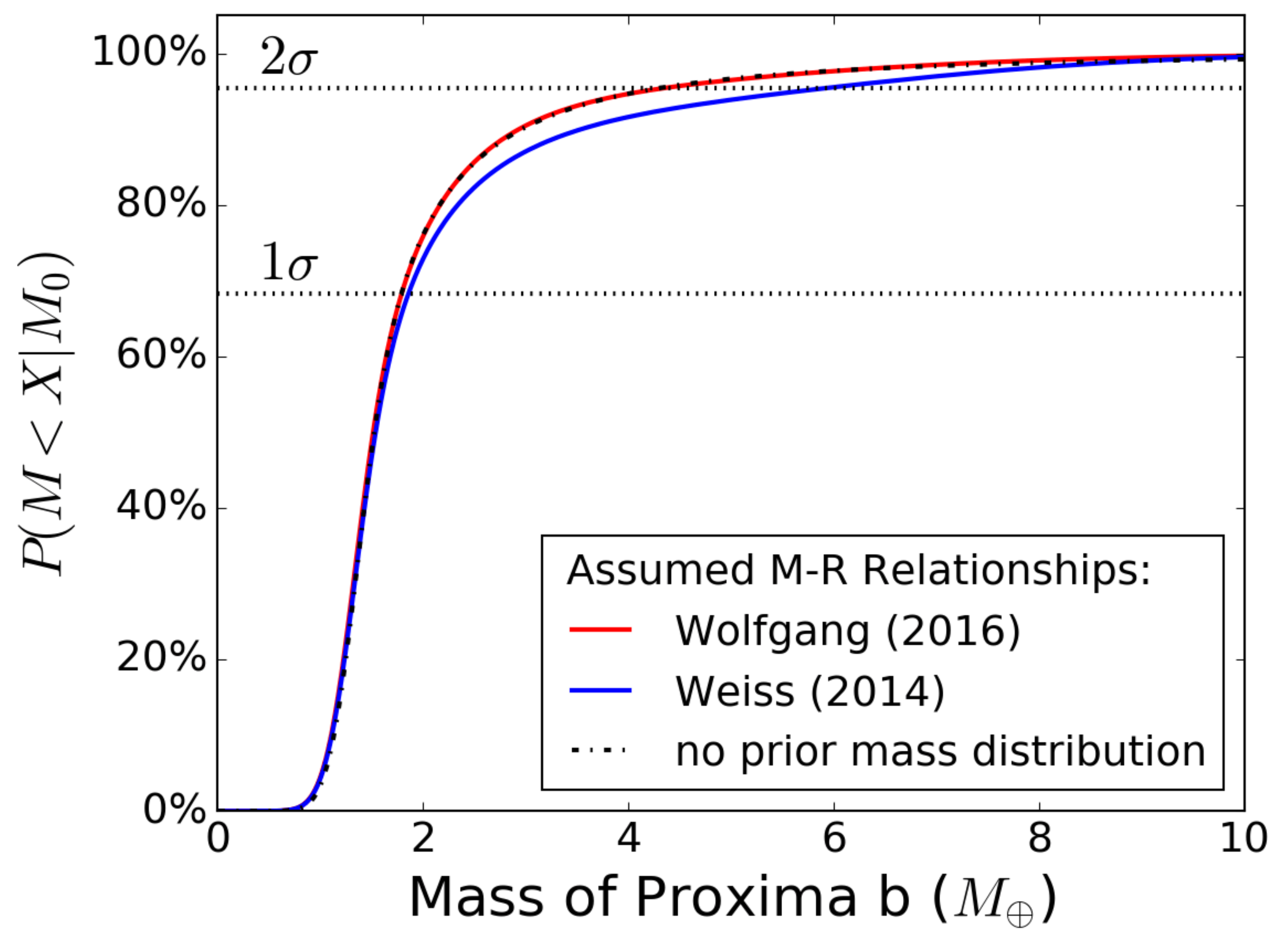}
\caption{The cumulative mass probability distribution for our simulated posterior mass distribution (solid) and assuming a flat prior $P(M)$ (dashdot). The dotted lines intersect 68\% and 95\% confidence upper limits on the mass. \label{fig:cumulative_pdf}}
\end{figure}

The prior and posterior mass probability distributions for Proxima~b are plotted in Figure \ref{fig:prior_posterior_pdf}. The shaded regions demonstrate the relative contributions of the populations at each mass. The prior distribution is valid for RV-detected planets around M dwarfs with intermediate periods ($2 < P < 50$ days) and radii ($0.7 < R < 4$ \re), while the posterior distribution can be taken as the mass probability distribution for Proxima~b.

For reference, we include the posterior distribution given \emph{no} prior constraints on the mass; that is, the distribution resulting from an isotropic $\sin(i)$ distribution and the measured $M_0$ with its uncertainty. We find that this nearly matches our result, since both $P(\sin(i))$ and $P(M)$ are bottom-heavy.

Figure \ref{fig:cumulative_pdf} shows the cumulative probability that $M < X$ for both of the considered M-R relationships (Section \ref{sec:mass-radius_relationships}) as well as for the case of no prior mass distribution. We find that there is little difference between the results for each M-R relationship.

\subsection{Escape velocity} \label{sec:escape}

In order to verify that sub-Neptune planets can maintain H$_2$ envelopes in the habitable zone, we compare the escape velocities of our simulated sub-Neptunes to the $\sim 8$ km/s cutoff for H$_2$ atmospheric escape (assuming an Earth-like stellar flux) defined by \citet{zahnle13}. In both the prior and posterior distributions of escape velocities, we find that fewer than 1\% of the sub-Neptunes have escape velocities below this threshold, with most having $v_e \gtrsim 15$ km/s. Therefore, we do not believe that Proxima b will be subject to significant atmospheric loss if it has a sub-Neptune composition.

\subsection{Composition} \label{sec:results_composition}

Table \ref{tab:cases} lists the sets of parameters for which we run the simulation, including the mass spread $dM$ for each composition and the central value ($R_\text{thresh}$) and width ($\Delta R$, if nonzero) of the transition region defined by Equation \ref{equation:rogers_linear}. The following results for each case are given: the probability $P_\text{rocky}$ that Proxima~b belongs to the `rocky' category of planets, i.e. that its density is \emph{consistent} with a fully iron and silicate composition, and the expectation values $\left<M\right>_\text{rocky}$ and $\left<R\right>_\text{rocky}$ of the mass and radius under the assumption that it belongs to this population.

Case A is most consistent with the previous work we have cited, so we take it as our primary result. In this case, there is a $\sim 90\%$ probability that Proxima~b belongs to the `rocky' population, with an $\sim 10\%$ likelihood that it belongs to the `sub-Neptune' population. In the case that it is rocky, the expectation values (and 95\% confidence intervals) for the mass and radius are $\left<M\right>_\text{rocky} = 1.63_{-0.72}^{+1.66}$ \me\ and $\left<R\right>_\text{rocky} = 1.07_{-0.31}^{+0.38}$ \re.

We investigate the effect of increasing (Case B) and decreasing (Case C) the mass spread for each composition, which results in lower and higher values of $P_\text{rocky}$, respectively. This results from low-radius ($R\sim 1.5$ \re), low-mass sub-Neptunes; when $dM$ is large, they can lie significantly below the M-R relation with masses between 1 and 2 \me, so that they are indistinguishable from the rocky planets in the mass domain.

In Cases D and E, we determine the effect of raising or lowering the threshold radius $R_\text{thresh}$ at which the rocky and sub-Neptune populations are split. A 0.2 \re\ offset in either direction, which encompasses most of the values suggested in the literature, results in a $\sim 5\%$ to $8\%$ shift in $P_\text{rocky}$, where higher threshold radii allow for more rocky planets and therefore a higher probability of a rocky composition. Furthermore, allowing for a non-zero width $\Delta R$ to the cutoff region allows sub-Neptunes to exist with lower radii and masses, thereby decreasing $P_\text{rocky}$.

In all cases, we find $P_\text{rocky}$ to be between $80\%$ and $95\%$ using the \citet{wolfgang16} M-R relationship, and we find similar values using the \citet{weiss14} relationship (e.g. $P_\text{rocky} = 90.7\%$ for Case A), so this result does not vary substantially over the range of reasonable values for the input parameters.

\section{Conclusions} \label{sec:conclusions}
By considering occurrence rates from the \emph{Kepler} mission and empirically derived M-R relationships, we derive a posterior probability distribution for the actual mass of Proxima~b. If the planet has a rocky composition, i.e. if it obeys the low-radius M-R relationship of \cite{wolfgang16}, then the expectation values of the mass and radius (with 95\% confidence intervals) are $\left<M\right>_\text{rocky} = 1.63_{-0.72}^{+1.66}$ \me\ and $\left<R\right>_\text{rocky} = 1.07_{-0.31}^{+0.38}$ \re.

In all of our simulations, we find a probability of 80\% to 95\% that Proxima~b belongs to the `rocky' population of planets defined in Section \ref{sec:compositions}. In our `best guess' scenario (Case A), this probability is 90\%. Critically, we note that we have assumed all planets with $f_\alpha = 1$ (according to the \cite{rogers15} criterion) are rocky planets, while in reality their density is only consistent with such a composition. With this in mind, it is safest to say that there is \emph{at least} a 10\% chance that Proxima~b has a sub-Neptune composition. If it is a sub-Neptune, then its surface gravity is high enough that it could maintain a thick hydrogen atmosphere.

For future theoretical work involving the habitability and detectability of Proxima~b, we advise caution regarding assumptions made about its mass or composition; if Proxima~b does possess a thick H/He envelope, then it is likely not habitable in the traditional sense. Even if the mass could be further constrained, sub-Neptunes have been measured with masses as low as $\sim 1$ \me, so the composition cannot be conclusively inferred from the mass alone.  Nevertheless, the rocky composition originally asserted by \citet{anglada16} remains the most likely possibility.

The results reported herein benefited from collaborations and/or information exchange within NASA's Nexus for Exoplanet System Science (NExSS) research coordination network sponsored by NASA's Science Mission Directorate. We thank Benjamin Rackham and Gijs Mulders for their constructive advice and insights, and the anonymous referee for their comments.

\floattable
\begin{deluxetable*}{cCccccCccc}
\tablecaption{Monte-Carlo Simulation Parameters and Results \label{tab:cases}}
\tablehead{
 \multicolumn{1}{c}{Case} && \multicolumn{4}{c}{Parameters} && \multicolumn{3}{c}{Results} \\ 
 \colhead{} && \colhead{dM (rocky)} & \colhead{dM (sub-Neptune)} & \colhead{$R_\text{thresh}$} & \colhead{$\Delta R$} && \colhead{$P_\text{rocky}$} & \colhead{$\left<M\right>_\text{rocky}$} & \colhead{$\left<R\right>_\text{rocky}$}
 }
\startdata
\textbf{Case A} && $ \mathbf{0.3\times M}$& \textbf{1.9 \meb} & \textbf{1.5 \reb} & \textbf{-} && $\mathbf{89.9\%}$ & $\mathbf{1.63_{-0.72}^{+1.66}}$ \meb & $\mathbf{1.07_{-0.31}^{+0.38}}$ \reb \\
Case B &&  $0.6\times M$ & 3.8 \me & 1.5 \re & - && $84.6\%$ &  $1.65_{-0.73}^{+1.95}$ \me & $1.03_{-0.36}^{+0.42}$ \re \\
Case C &&  $0.1\times M$ & 0.7 \me & 1.5 \re & - && $93.6\%$ &  $1.65_{-0.73}^{+1.52}$ \me & $1.06_{-0.24}^{+0.36}$ \re \\
Case D &&  $0.3\times M$ & 1.9 \me & 1.7 \re & - && $94.6\%$ &  $1.71_{-0.79}^{+2.13}$ \me & $1.10_{-0.33}^{+0.50}$ \re \\
Case E &&  $0.3\times M$ & 1.9 \me & 1.3 \re & - && $81.6\%$ &  $1.52_{-0.62}^{+1.15}$ \me & $1.02_{-0.27}^{+0.26}$ \re \\
Case F &&  $0.3\times M$& 1.9 \me & 1.5 \re & 1.2 \re && $84.8\%$ &  $1.64_{-0.73}^{+1.99}$ \me & $1.06_{-0.30}^{+0.53}$ \re \\
Case G &&  $0.6\times M$& 3.8 \me & 1.5 \re & 1.2 \re && $81.1\%$ &  $1.65_{-0.75}^{+2.13}$ \me & $1.02_{-0.36}^{+0.63}$ \re \\
\enddata
\tablecomments{The resulting values of $P_\text{rocky}$, $\left<M\right>_\text{rocky}$, and $\left<R\right>_\text{rocky}$ for different mass spreads $dM$ and compositional parameters $R_\text{thresh}$ and $\Delta R$. The expectation values are reported with 95\% confidence intervals.}
\end{deluxetable*}





\begin{thebibliography}{}
\expandafter\ifx\csname natexlab\endcsname\relax\def\natexlab#1{#1}\fi

\bibitem[{{Anglada-Escud{\'e}} {et~al.}(2016){Anglada-Escud{\'e}}, {Amado},
  {Barnes}, {Berdi{\~n}as}, {Butler}, {Coleman}, {de La Cueva}, {Dreizler},
  {Endl}, {Giesers}, {Jeffers}, {Jenkins}, {Jones}, {Kiraga}, {K{\"u}rster},
  {L{\'o}pez-Gonz{\'a}lez}, {Marvin}, {Morales}, {Morin}, {Nelson}, {Ortiz},
  {Ofir}, {Paardekooper}, {Reiners}, {Rodr{\'{\i}}guez},
  {Rodr{\'{\i}}guez-L{\'o}pez}, {Sarmiento}, {Strachan}, {Tsapras}, {Tuomi}, \&
  {Zechmeister}}]{anglada16}
{Anglada-Escud{\'e}}, G., {Amado}, P.~J., {Barnes}, J., {et~al.} 2016, \nat,
  536, 437

\bibitem[{{Barnes} {et~al.}(2016){Barnes}, {Deitrick}, {Luger}, {Driscoll},
  {Quinn}, {Fleming}, {Guyer}, {McDonald}, {Meadows}, {Arney}, {Crisp},
  {Domagal-Goldman}, {Lincowski}, {Lustig-Yaeger}, \&
  {Schwieterman}}]{barnes16}
{Barnes}, R., {Deitrick}, R., {Luger}, R., {et~al.} 2016, ArXiv e-prints,
  arXiv:1608.06919

\bibitem[{{Belikov} {et~al.}(2015){Belikov}, {Bendek}, {Thomas}, {Males}, \&
  {Lozi}}]{belikov15}
{Belikov}, R., {Bendek}, E., {Thomas}, S., {Males}, J., \& {Lozi}, J. 2015, in
  \procspie, Vol. 9605, Techniques and Instrumentation for Detection of
  Exoplanets VII, 960517

\bibitem[{{Brugger} {et~al.}(2016){Brugger}, {Mousis}, {Deleuil}, \&
  {Lunine}}]{brugger16}
{Brugger}, B., {Mousis}, O., {Deleuil}, M., \& {Lunine}, J.~I. 2016, \apjl,
  831, L16

\bibitem[Coleman et al.(2017)]{coleman17} Coleman, G.~A.~L., Nelson, R.~P., Paardekooper, S.~J., et al.\ 2017, \mnras,  

\bibitem[{{Goldblatt}(2016)}]{goldblatt16}
{Goldblatt}, C. 2016, ArXiv e-prints, arXiv:1608.07263

\bibitem[{{Ho} \& {Turner}(2011)}]{ho11}
{Ho}, S., \& {Turner}, E.~L. 2011, \apj, 739, 26

\bibitem[Kervella et al.(2017)]{kervella17} Kervella, P., Th{\'e}venin, F., \& Lovis, C.\ 2017, \aap, 598, L7 

\bibitem[{{Lovis} {et~al.}(2016){Lovis}, {Snellen}, {Mouillet}, {Pepe},
  {Wildi}, {Astudillo-Defru}, {Beuzit}, {Bonfils}, {Cheetham}, {Conod},
  {Delfosse}, {Ehrenreich}, {Figueira}, {Forveille}, {Martins}, {Quanz},
  {Santos}, {Schmid}, {S{\'e}gransan}, \& {Udry}}]{lovis16}
{Lovis}, C., {Snellen}, I., {Mouillet}, D., {et~al.} 2016, ArXiv e-prints,
  arXiv:1609.03082

\bibitem[{{Luger} {et~al.}(2016){Luger}, {Lustig-Yaeger}, {Fleming}, {Tilley},
  {Agol}, {Meadows}, {Deitrick}, \& {Barnes}}]{luger16}
{Luger}, R., {Lustig-Yaeger}, J., {Fleming}, D.~P., {et~al.} 2016, ArXiv
  e-prints, arXiv:1609.09075

\bibitem[{{Mulders} {et~al.}(2015){Mulders}, {Pascucci}, \& {Apai}}]{mulders15}
{Mulders}, G.~D., {Pascucci}, I., \& {Apai}, D. 2015, \apj, 814, 130

\bibitem[Owen \& Mohanty(2016)]{owen16} Owen, J.~E., \& Mohanty, S.\ 2016, \mnras, 459, 4088 


\bibitem[Ribas et al.(2016)]{ribas16} Ribas, I., Bolmont, E., Selsis, F., et al.\ 2016, \aap, 596, A111 

\bibitem[{{Rogers}(2015)}]{rogers15}
{Rogers}, L.~A. 2015, \apj, 801, 41

\bibitem[{{Schwieterman} {et~al.}(2016){Schwieterman}, {Meadows},
  {Domagal-Goldman}, {Deming}, {Arney}, {Luger}, {Harman}, {Misra}, \&
  {Barnes}}]{schwieterman16}
{Schwieterman}, E.~W., {Meadows}, V.~S., {Domagal-Goldman}, S.~D., {et~al.}
  2016, \apjl, 819, L13

\bibitem[Turbet et al.(2016)]{turbet16} Turbet, M., Leconte, J., Selsis, F., et al.\ 2016, \aap, 596, A112 

\bibitem[{{Weiss} \& {Marcy}(2014)}]{weiss14}
{Weiss}, L.~M., \& {Marcy}, G.~W. 2014, \apjl, 783, L6

\bibitem[{{Wolfgang} {et~al.}(2016){Wolfgang}, {Rogers}, \&
  {Ford}}]{wolfgang16}
{Wolfgang}, A., {Rogers}, L.~A., \& {Ford}, E.~B. 2016, \apj, 825, 19

\bibitem[Zahnle \& Catling(2013)]{zahnle13} Zahnle, K.~J., \& Catling, D.~C.\ 2013, Lunar and Planetary Science Conference, 44, 2787 

\end{thebibliography}




\end{document}